\documentclass[aps,pre,twocolumn,groupedaddress,floatfix]{revtex4-1}
\usepackage{graphicx}
\usepackage{color}
\usepackage{amsmath}
\usepackage{array}

\newcommand{\critDrop}{\langle\tilde{\eta}\rangle_{T_c}}
\newcommand{\tauW}{\tau_{\rm W}}
\newcommand{\Ising}{^{\rm Is}}
\newcommand{\tauWI}{\tau_{\rm W}\Ising}
\newcommand{\TI}{T\Ising}

\newcommand{\rhoL}{\rho_{\rm L}}
\newcommand{\ND}{N_{\rm D}}
\newcommand{\NGas}{N_{\rm G}}
\newcommand{\VD}{V_{\rm D}}
\newcommand{\kB}{k_{\rm B}}
\newcommand{\mykappa}{\hat{\kappa}}

\begin{document}

\title{
  Exploring different regimes in finite-size scaling of the droplet
  condensation-evaporation transition
}
\author{Johannes Zierenberg}
\email[]{johannes.zierenberg@itp.uni-leipzig.de}
\affiliation{Institut f\"ur Theoretische Physik, 
             Universit\"at Leipzig, 
             Postfach 100\,920, 
             D-04009 Leipzig, 
             Germany
           }
\author{Wolfhard Janke}
\email[]{wolfhard.janke@itp.uni-leipzig.de}
\affiliation{Institut f\"ur Theoretische Physik, 
             Universit\"at Leipzig, 
             Postfach 100\,920, 
             D-04009 Leipzig, 
             Germany
           }

\date{\today}

\begin{abstract}
  We present a finite-size scaling analysis of the droplet
  condensation-evaporation transition of a lattice gas (in two and three
  dimensions) and a Lennard-Jones gas (in three dimensions) at fixed density.
  Parallel multicanonical simulations allow sampling of the required system
  sizes with precise equilibrium estimates.  In the limit of large systems, we
  verify the theoretical leading-order scaling prediction for both the
  transition temperature and the finite-size rounding. In addition, we present
  an emerging intermediate scaling regime, consistent in all considered cases
  and with similar recent observations for polymer aggregation.  While the
  intermediate regime locally may show a different effective scaling, we show
  that it is a gradual crossover to the large-system scaling behavior by
  including empirical higher-order corrections. This implies that care has to be
  taken when considering scaling ranges, possibly leading to completely wrong
  predictions for the thermodynamic limit.  In this study, we consider a
  crossing of the phase boundary orthogonal to the usual fixed temperature
  studies. We show that this is an equivalent approach and, under certain
  conditions, may show smaller finite-size corrections.
\end{abstract}


\maketitle


\section{Introduction}
The limit of large systems, or the thermodynamic limit, is of general interest
when studying finite systems. It is this limit, that is commonly accessible to
experiments. Recent developments in experimental and simulation techniques open
the door to mesoscopic length scales~\cite{MuellerPlathe2002}.  Here, general
concepts of finite-size scaling may be tested and verified.  Considering, for
example, the equilibrium properties of several homopolymers, the aggregation
transition temperature was shown to exhibit systematic finite-size effects that
deviate from the expected behavior for particle gas
systems~\cite{Zierenberg2014JCP}. It was argued that the observed scaling is
valid for intermediate system sizes, where the aggregate includes most of the
polymers, but in the limit of increasing polymer number the particle picture
should be recovered. On the other hand, it is a reasonable assumption that the
intermediate regime should be also apparent for particle gas condensation, which
will be a main focus of the present paper.

First-order phase transitions in spin systems are usually separating homogeneous
phases, which leads in general to finite-size scaling corrections on the order
of the inverse system volume. Exceptions occur for example if the
low-temperature phase is exponentially degenerated~\cite{Mueller2014}. The
general situation changes when we consider a phase transition between a
homogeneous phase and a mixed phase -- as in the case of the droplet
condensation-evaporation transition separating a supersaturated gas phase and a
mixed phase consisting of a single droplet in equilibrium with surrounding
vapor. There exists a large amount of theoretical literature on this topic about
the leading-order scaling behavior of large
systems~\cite{Biskup2002,Neuhaus2003,Binder2003} with origins already in the
80s~\cite{Binder1980}. Usually, a system at fixed temperature is considered,
estimating the transition density at which a single macroscopic droplet forms.
In the limit of very large systems, this transition density is supposed to have
finite-size corrections of the order $V^{-1/(d+1)}$, while the rounding of the
transition should scale with $V^{-d/(d+1)}$. The former was numerically verified
to leading-order at fixed temperature for a lattice gas~\cite{Neuhaus2003,
Nussbaumer2006, Nussbaumer2008, Zierenberg2014JPCS}.  Several studies of a
three-dimensional Lennard-Jones gas~\cite{Binder2004,MacDowell2006,Binder2009}
also verified the finite-size corrections of the transition density. For a fixed
density, the scaling of the transition temperature was demonstrated for the two-
and three-dimensional lattice gas~\cite{Malakis2007}, while the scaling exponent
of the finite-size rounding could not be verified. There exist complementary
microcanonical studies of the Ising model at fixed density (or
magnetization)~\cite{Pleimling2001, Pleimling2009}, demonstrating that the
occurring transition shows signatures of a first-order transition.

The present study aims to fill the gap with respect to the finite-size scaling
of the droplet condensation-evaporation transition at fixed density and the
gradual crossover from an effective intermediate regime to the asymptotic
large-system limit.
In order to draw general conclusions, we consider a lattice-gas model in two
and three dimensions as well as a Lennard-Jones gas in three dimensions. We
will show that the proposed scaling behavior of the transition temperature and
rounding is numerically recaptured in the limit of large (but achievable)
system sizes for all dimensions and models considered. Where possible, we will
directly compare to analytic solutions or low-temperature series expansions.
In addition, we discuss an emerging intermediate scaling regime, best visible
in the rounding of the transition and consistent with recent observations of
finite-size effects in polymer aggregation~\cite{Zierenberg2014JCP}. This
regime with effective local scaling behavior and the gradual crossover to the
large-system regime may be described by the finite-size scaling behavior when
including empirical higher-order correction terms.%

The remaining part of this paper is organized as follows. In
Sec.~\ref{sec_theory} we briefly recapture the main leading-order results from
the literature at fixed temperature and convert them to the scenario of fixed
density. After describing our models and methods in Sec.~\ref{sec_model}, we
will present our results in Sec.~\ref{sec_results} and finish with the
conclusions in Sec.~\ref{sec_conclusions}.

\section{Theory}
\label{sec_theory}
\begin{figure}
  \centering
  \includegraphics{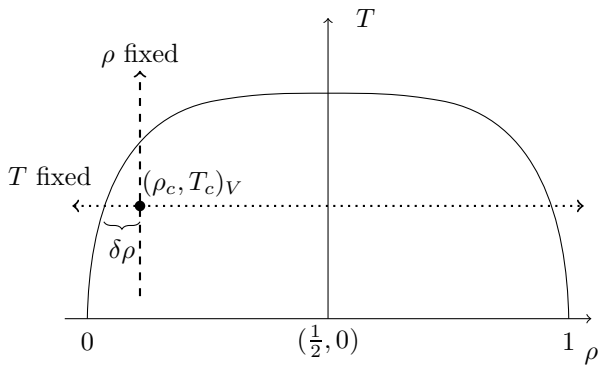}
  \caption{Sketch of the infinite system-size transition (solid line) together
    with the finite-size scaling directions in either the density ($T$ fixed) or
    temperature ($\rho$ fixed). The transition line may be understood as
    $T_0(\rho)$ or similarly $\rho_0(T)$. At the crossing point of both schemes
    a finite system of size $V$ may be constructed for which
    $(\rho,T)=(\rho_c,T_c)_V$ corresponds to the finite-size
    condensation-evaporation transition. 
  }
  \label{fig_DensityToTemperature}
\end{figure}
We consider a $d$-dimensional liquid-vapor system of $N$ particles in a
(periodic) box of volume $V$, where for sufficiently dilute systems the
condensation-evaporation transition separates a homogeneous supersaturated gas
phase from a mixed phase of a droplet in equilibrium with surrounding bulk gas.
At fixed density, this occurs at the condensation-evaporation transition
temperature $T_c$.  However, the usual description of this problem is in the
language of a grand-canonical scheme, considering a fixed temperature and
variable density (or in case of a lattice gas equivalently the magnetization of
the Ising model) with the finite-size transition density $\rho_c$. This is
orthogonal to the scenario of a fixed density with variable temperature.
Figure~\ref{fig_DensityToTemperature} shows a sketch of the infinite-size phase
boundary $T_0(\rho)$ or $\rho_0(T)$ and the different scenarios crossing it for
finite system sizes. It needs to be mentioned that both schemes are working in
the canonical ensemble for each point $(\rho,T)$.  
So in fact, any finite-size transition point $(\rho_c,T_c)_V$ belongs to one
fixed-$T$ and one fixed-$\rho$ scheme, simultaneously. The same holds for any
canonical function $f(\rho,T)$.  Thus, the orthogonal crossing schemes are
equivalent and we may translate a functional dependence $f(\rho,T)V^\alpha=1$
from one scheme to another by a Taylor series expansion. Then, expanding around
some $T^*$ yields
\begin{equation}
  V^{-\alpha}=f(\rho,T^*)+f'(\rho,T^*)(T-T^*) + ...~,
\end{equation}
which may be solved for $T$. The remaining task is to identify and evaluate
suitable functional dependencies.

Using this procedure, we will recapture for a fixed density $\rho=N/V$ the
finite-size correction to the transition temperature \mbox{$T_c-T_0\propto
N^{-1/(d+1)}$} and the scaling of the finite-size rounding (or width of the
finite-system transition region) $\Delta T=T-T_c\propto N^{-d/(d+1)}$.  This is
directly related to the linear extension of the droplet $R\propto N^{1/(d+1)}$,
which becomes the relevant length scale for large system sizes.%

%
\subsection{Finite-size scaling of the droplet condensation-evaporation transition}
\label{sec_fss_temp}
Biskup et al.~\cite{Biskup2002} showed for a $d$-dimensional liquid-vapor system
in equilibrium a vanishing probability of intermediate-sized droplets. This was
proven rigorously for the lattice-gas interpretation of the 2D Ising model and
justifies the restriction of the discussion to two relevant phases: the
homogeneous gas phase and the mixed phase of a single macroscopic droplet in
equilibrium with a bulk gas surrounding it. At coexistence both phases are
equally probable and in general (finite volume) the dominance of either phase is
the solution of a saddle point problem (see also Ref.~\cite{Binder2003}).
Choosing a \emph{fixed temperature} $T$, we consider a system at supersaturated
density $\rho=N/V$ with particle excess $\delta N = N-N_0 = \delta\rho V$, where
$\rho_0=N_0/V$ is the equilibrium density of the infinite-system gas phase at
this temperature. In fact, the particle excess is temperature dependent, see
Fig.~\ref{fig_DensityToTemperature}. A fraction $\lambda\in[0,1]$ of this
particle excess is considered to be in the single macroscopic droplet, i.e.,
$\delta\ND=\lambda\delta N$, while the remaining excess is in the gas phase such
that $\delta N=\delta\ND + \delta\NGas$. Up to higher-order corrections from
surface and density fluctuations as well as translational entropy of the droplet
motion, Biskup et al.~\cite{Biskup2002} introduced a rescaled
(size-independent) density
\begin{equation}
  \Delta = \frac{(\rhoL-\rho_0)^{\frac{d-1}{d}}}{2\hat{\kappa} \tauW} \frac{(\delta
  N)^{\frac{d+1}{d}}}{V},
\label{eq_Delta}
\end{equation}
with the infinite-system liquid and gas densities $\rhoL$ and $\rho_0$, the
Wulff-shape surface free-energy per unit volume $\tauW$ and the (reduced)
isothermal compressibility $\hat{\kappa}$, see also
Ref.~\cite{Zierenberg2014JPCS}. Here, only $\delta N$ is variable and the
remaining parameters are constants encoding all system-specific details, like
the droplet shape. In the limit of large system sizes, they calculate the
fraction of particles in the largest droplet $\lambda(\Delta)$ as a
``universal'' function of $\Delta$. At the transition density
$\Delta_c=\frac{1}{d}\left(\frac{d+1}{2}\right)^{(d+1)/d}$ the system changes
between the vapor phase ($\lambda=0$) and the mixed phase
(\mbox{$\lambda_c=2/(d+1)$}).
 
Now consider a \emph{fixed density} $\rho$ with varying temperature $T$. Then
the system-specific constants become functions of the temperature, namely
$\rho_i(T)$, $\mykappa(T)$, $\tauW(T)$ and \mbox{$\delta N(T)=\left(\rho
-\rho_0(T)\right)V$}. The temperature-dependent particle excess may be
understood by the fact that for decreasing temperature (fixed density) the
difference $\delta\rho$ increases and with it the supersaturation, see also
Fig.~\ref{fig_DensityToTemperature}. We may rewrite Eq.~\eqref{eq_Delta} as 
\begin{equation}
  \Delta^{\frac{d}{d+1}} V^{-\frac{1}{d+1}} = f(\rho,T), 
\label{eq_koteckySizeDep}
\end{equation}
where we identify 
\begin{equation}
  f(\rho,T) = \frac{\rho -\rho_0(T)}{\rhoL(T)-\rho_0(T)}
  \left(\frac{(\rhoL(T)-\rho_0(T))^2}{2\mykappa(T)\tauW(T)}\right)^{\frac{d}{d+1}}.
\end{equation}
At the condensation-evaporation transition, $\Delta=\Delta_c$ is constant and
the left-hand side of Eq.~\eqref{eq_koteckySizeDep} is depending only on the
system size $V$.  Then, for a fixed finite system size, a suitable combination
of $T$ and $\rho$ solves Eq.~\eqref{eq_koteckySizeDep} yielding the finite-size
transition point at $(\rho,T)=(\rho_c,T_c)_V$ in
Fig.~\ref{fig_DensityToTemperature}. This transition point may be obtained
either numerically exact or by a Taylor expansion. Keeping $\rho=N/V$ constant,
we proceed by expanding $f(\rho,T)$ in Eq.~\eqref{eq_koteckySizeDep} around the
infinite-system transition temperature $T_0$, where $\rho_0(T_0)=\rho$ and thus
$f(\rho,T_0)=0$. Solving for the finite-size transition temperature
$T=T_c$ yields
\begin{equation}
  T_c = T_0 + \frac{\Delta_c^{\frac{d}{d+1}}}{f'(\rho,T_0)}V^{-\frac{1}{d+1}}
  + \mathcal{O}\left(V^{-\frac{2}{d+1}}\right).
  \label{eq_kotecky_expansion}
\end{equation}
In terms of the number of particles this means to first order
\begin{equation}
  T_c-T_0\propto N^{-\frac{1}{d+1}}.
\end{equation}
We need to emphasize that Eq.~\eqref{eq_koteckySizeDep} is only the leading-order
result. Numeric tests at fixed temperature already showed apparent higher-order
corrections~\cite{Nussbaumer2006, Nussbaumer2008, Zierenberg2014JPCS}.

For the Ising model, we can go a little further and evaluate the leading
finite-size scaling explicitly. To this end, we relate the magnetization and the
density via $m=1-2\rho$ and identify the spontaneous magnetization
$m_0=1-2\rho_0$ as well as, from the symmetry of the Ising model,
$\rhoL=1-\rho_0$. In addition, we find the magnetic susceptibility
$\chi=\mykappa$ and $\tauWI=4\tau_W$, due to the shift in the energy scale
when exploiting the equivalence of the Ising model and the lattice gas
model~\cite{Zierenberg2014JPCS}. Thus, we obtain
\begin{equation}
  f(m,\TI) = \frac{1}{2}\left(1-\frac{m}{m_0(\TI)}\right)
  \left(\frac{2m_0(\TI)^2}{\chi(\TI)\tauWI(\TI)}\right)^{\frac{d}{d+1}},
\label{eq_koteckyIsing}
\end{equation}
with $\TI=4T$ when rewriting the Hamiltonian according to the definition in
Sec.~\ref{sec_model} and Ref.~\cite{Zierenberg2014JPCS}.

In two dimensions, the involved quantities are known analytically or up to
arbitrary precision: $m_0$ is described by the Onsager-Yang
equation~\cite{Onsager}, $\chi$ is obtained from sufficiently long series
expansions~\cite{Orrick2001,Nickelchi2,chi2new} and $\tauWI=2\sqrt{W}$ can be
obtained from the volume of the Wulff plot $W$~\cite{Zia1990}. For a collection
of equations we refer to Ref.~\cite{Nussbaumer2008}. We numerically evaluated
Eq.~\eqref{eq_koteckyIsing} in Eq.~\eqref{eq_koteckySizeDep}, fixing the
density and volume and solving with a bisection algorithm for the corresponding
transition temperature. We will compare this result to the numerical
finite-size results later in Fig.~\ref{fig_results_fss_temp}~(top) and refer to
it as the \emph{full solution} of Eq.~\eqref{eq_koteckySizeDep}.

In three dimensions, we may make use of low-temperature series expansions of the
spontaneous magnetization~\cite{3DLowTempM}. This allows us to estimate the
infinite-system transition temperature $T_0$ by solving
\mbox{$\rho-\rho_0(T)=0$}. We will compare these results to the finite-size
scaling fits of our numerical data and refer to it as the solution from
\emph{low-temperature series expansion} in
Fig.~\ref{fig_results_fss_temp}~(center).

The analytic results may be used to recapture the scaling of the droplet size at
condensation, where the fraction $\lambda_c$ will be in the largest droplet. For
the volume of the droplet at condensation it follows $\delta\VD = (\rho_L
-\rho_0)^{-1} \lambda_c\delta N_c$~\cite{Biskup2002}, where the total particle
excess at condensation $\delta N_c$ encodes the shape of the droplet (see
Eq.~\eqref{eq_Delta}). In general, the volume of an ideal droplet may be
expressed by $\delta\VD=S_dR^d$, where $S_d$ is a geometric shape factor that
allows to describe both spherical and cubic droplets, where the latter may occur
in lattice systems below the roughening transition~\cite{Zierenberg2014JPCS}.
Equating both droplet volumes, inserting $\delta N_c$ from Eq.~\eqref{eq_Delta},
and solving for the radius yields
\begin{equation}
  R = \left(S_d^{-1} \delta V_D\right)^{\frac{1}{d}} \propto V^{\frac{1}{d+1}},
\label{eq_criticalRadius}
\end{equation}
consistent with results in the literature~\cite{Binder1980, Binder2003}. For
fixed density, this is equivalent to $R\propto N^{1/(d+1)}$. Moreover, the leading
finite-size scaling corrections in Eq.~\eqref{eq_kotecky_expansion} may be expressed
in powers of $R^{-1}$, i.e., $T_c-T_0\propto R^{-1}$. 

%
\subsection{Rounding of the droplet condensation-evaporation transition}
\label{sec_fss_rounding}

In analogy, one may argue that the rounding of the transition at fixed density
should scale with the system size in the same way as the rounding of the
transition density (or magnetization) at fixed temperature. The latter was derived by
Binder~\cite{Binder2003} using a phenomenological theory. He predicted a
finite-size rounding proportional to $V^{-d/(d+1)}$. Starting from a two-state
approximation the discussion may be reduced to a homogeneous and an
inhomogeneous phase. Then, the rounding of the transition may be related to a
characteristic width $\beta\Delta F$ of order unity, where $\Delta F$ is the
free-energy difference between both phases. 

Expanding the free-energy difference $\beta\Delta F$ around the finite-size
transition temperature $T_c$, recalling that $(\partial/\partial\beta) \beta F
= E$, yields
\begin{equation}
  \beta\Delta F = \left.(\beta \Delta F)\right|_{T_c} - \left.\left(\frac{1}{\kB
  T^2}\Delta E\right)\right|_{T_c} (T-T_c) + ...~.
  \label{eq_freeEnergyExpanded}
\end{equation}
The free-energy difference vanishes at $T_c$ in the limit of large system
sizes, considering that both phases contribute with equal probability. 
In the
gas phase, the particles may be considered non-interacting. Thus, the energy
difference is dominated by the droplet energy, which depends on the droplet
volume $\propto R^{d}$. At $T_c$, Eq.~\eqref{eq_criticalRadius} relates the
droplet radius $R$ to the system volume and the energy difference may be
approximated to leading order as $\Delta E \sim V^{d/(d+1)}$. 
The finite-size corrections from the transition temperature appear merely as
corrections to the energy difference, such that
Eq.~\eqref{eq_freeEnergyExpanded} simplifies to $\beta\Delta F \sim
(V^{d/(d+1)}/\kB T_0^2)~\Delta T$ in the limit of large system sizes. The
condition $|\beta\Delta F|\sim 1$ yields to leading order the rounding width
$\Delta T\propto V^{-d/(d+1)}$ and in terms of the particle number
\begin{equation}
  \Delta T \propto N^{-\frac{d}{d+1}}.
\end{equation}
Notice that the radius of the droplet $R$ as relevant length scale was used in the
argumentation such that for large system sizes the rounding may be identified as
$\Delta T=T-T_c\propto R^{-d}$.
%

\section{Models and Methods}
\label{sec_model}

\begin{figure}
  \centering
  \includegraphics{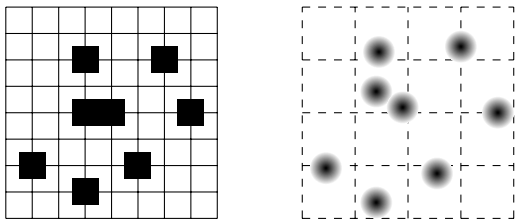}
  \caption{Sketch of the particle gas models under investigation: The discrete
    lattice gas model (left) in two and three dimensions and the continuous
    Lennard-Jones gas model with domain decomposition (right) in three dimensions.
  }
  \label{fig_model}
\end{figure}

In order to investigate the universal aspects of the condensation-evaporation
transition, we employ two different particle gas models, for a sketch see
Fig.~\ref{fig_model}. Bridging the gap to analytic solutions~\cite{Biskup2002},
we consider a lattice gas model equivalent to the Ising model at fixed
magnetization~\cite{Pleimling2001,Nussbaumer2006}. We add to our investigation a
Lennard-Jones gas model, which --- in contrast to the lattice gas --- is not
symmetric with respect to particle-hole exchange (see also
\cite{Binder2004,Binder2009}). The boundary conditions are periodic, which is
common for a finite-size scaling analysis. That way, non-intended interactions
with the boundaries are avoided. For both models, we consider a finite
interaction range smaller than the linear extension of the system, which safely
allows the use of periodic boundary conditions.
\begin{figure*}
  \centering
  \includegraphics[width=0.35\textwidth]{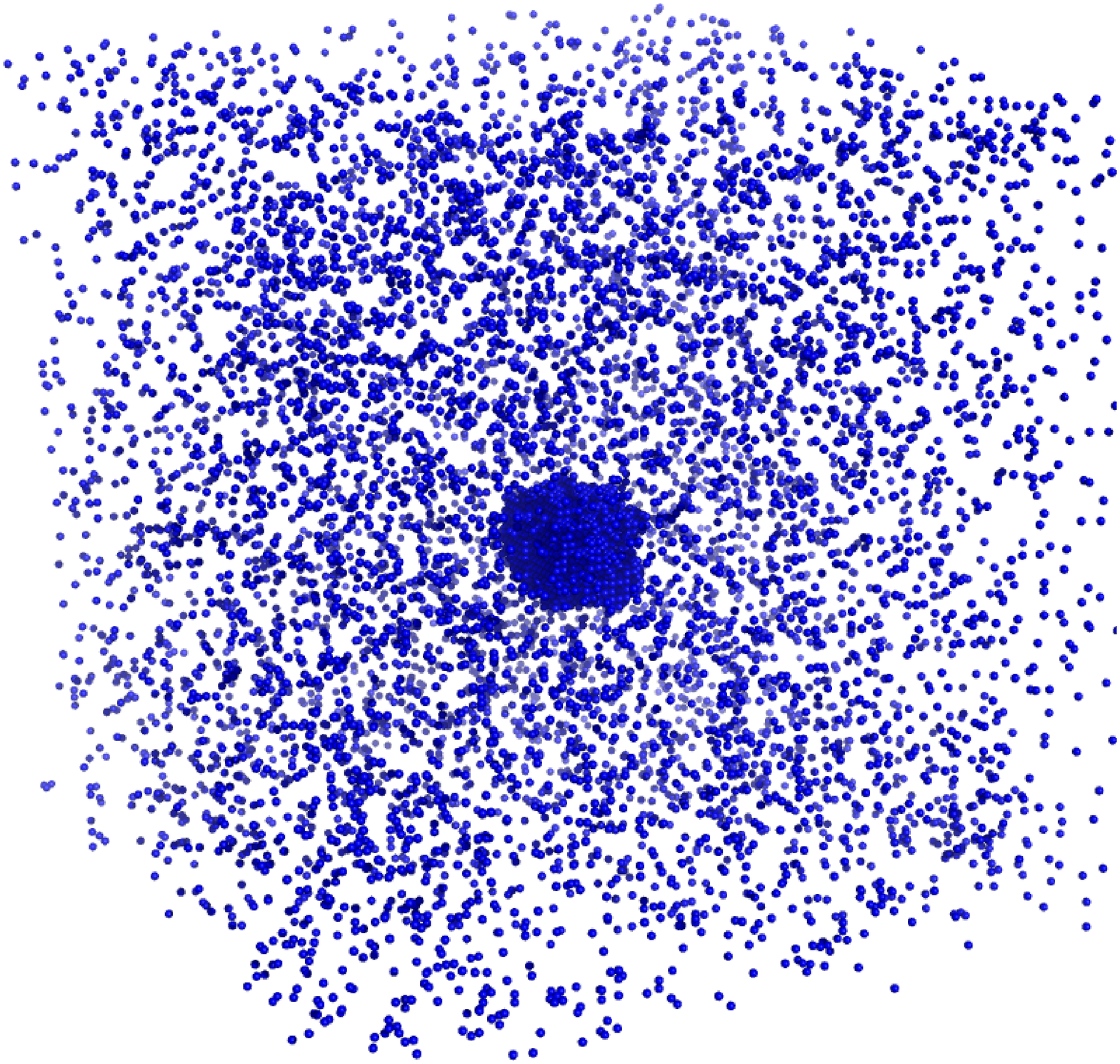}
  \hspace{5em}
  \includegraphics[width=0.35\textwidth]{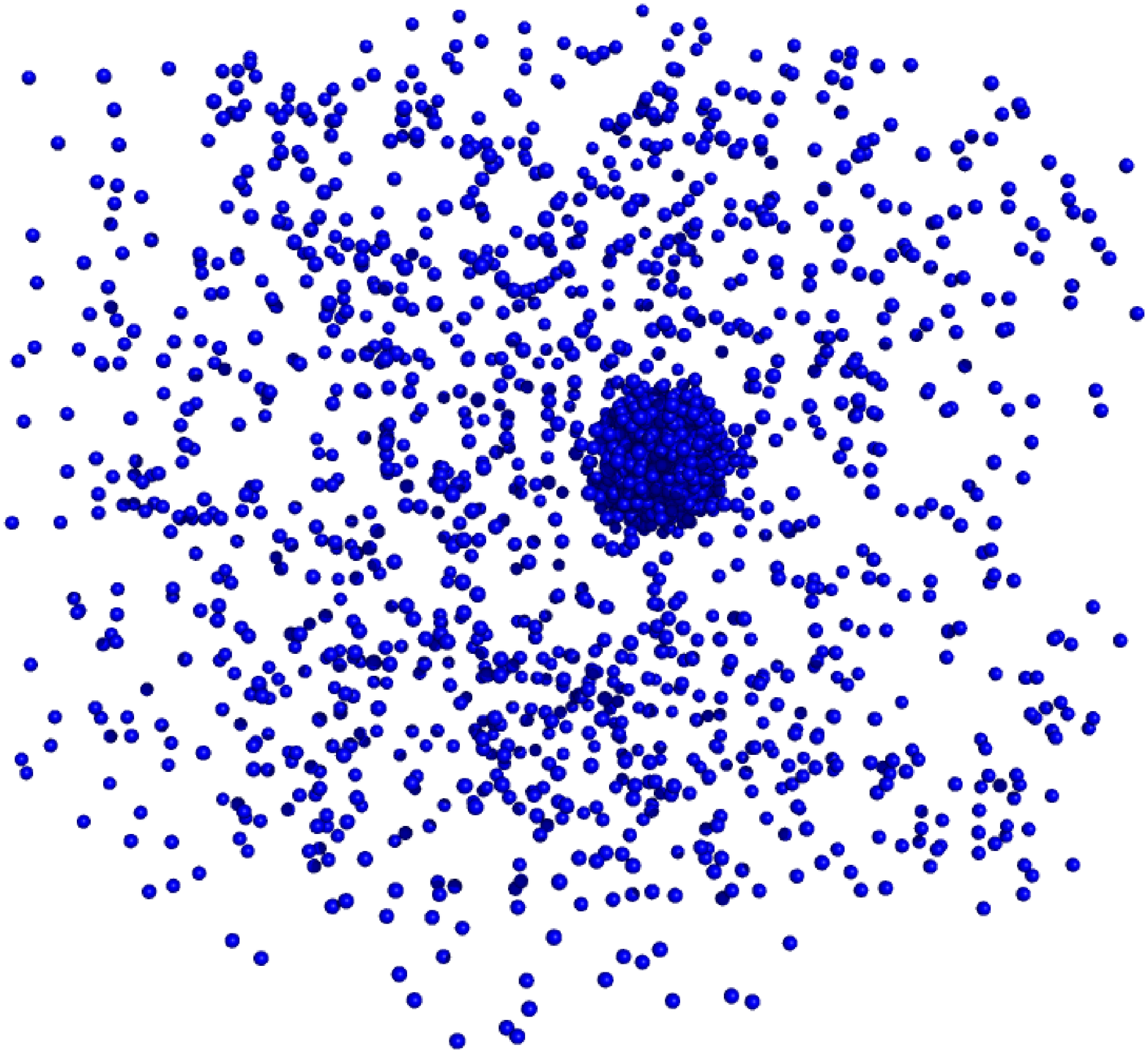}
  \caption{(Color online)
    Example of equilibrium droplets at coexistence from Metropolis simulations
    for the 3D lattice gas with $N=10\,000$ (left) and the 3D Lennard-Jones gas
    with $N=2\,000$ (right).
  }\label{fig_equilDroplet}
\end{figure*}
%

%
\subsection{Lattice gas}
In the discrete lattice gas (DLG) model, particles are described by occupied
sites on a two-dimensional (2D) square or three-dimensional (3D) cubic lattice
of size $V=L^d$. The interaction is purely short-range, with attraction between
nearest neighbors and hard-core repulsion by the condition that each site can be
occupied by only one particle.  Introducing a state-variable for each site,
$n_i\in\{0,1\}$, the Hamiltonian may be written as
\begin{equation}
  \mathcal{H}_{\rm DLG} = -\sum_{\langle i,j\rangle}n_i n_j,
\end{equation}
equivalent to the Ising model ($s_i\in\{\pm 1\}$) at fixed magnetization if
$n_i=\frac{1}{2}\left(s_i+1\right)$. Here, the density $\rho=N/V$ can only be
adjusted approximately by occupying an integer number of sites $N=\rho L^d$ .

%
\subsection{Lennard-Jones gas}
\label{sec_model_LJ}
The Lennard-Jones gas (LJG) model is a continuous particle gas model, where in
principle all particles interact with each other via the potential
\begin{equation}
  V_{\rm LJ}(r_{ij}) = 4\epsilon\left[\left(\frac{\sigma}{r_{ij}}\right)^{12} 
  - \left(\frac{\sigma}{r_{ij}}\right)^6\right],
\end{equation}
with $\epsilon=1$ and $\sigma=2^{-1/6}$ such that the potential minimum is
at $r_{\rm min} = 1$. 
Being consistent with literature and reducing the computational demand, we
introduce a cutoff radius $r_c=2.5\sigma$ above which the potential is zero.
The remaining potential is then shifted by $V_{\rm LJ}(r_c)$ in order to be
continuous: 
\begin{equation}
  V^*_{\rm LJ}(r) = 
  \begin{cases}
    V_{\rm LJ}(r)-V_{\rm LJ}(r_c) & r < r_c \\
    0                             & \mathrm{else}
  \end{cases}.
\end{equation}
That way, we may use a domain decomposition, only calculating interactions with
particles in neighboring domains. This reduces the computational demand
especially in the gas phase. The Hamiltonian is given by
\begin{equation}
  \mathcal{H}_{\rm LJG} = \frac{1}{2}\sum_{i\neq j} V_{\rm LJ}^*(r_{ij}).
\end{equation}
Here, we consider only the three-dimensional case of particles in a cubic box of
size $V=L^3$ with periodic boundary conditions and $L=(N/\rho)^{1/3}$.

For the present study, we considered the Lennard-Jones gas with the same density
as for the lattice gas. The Lennard-Jones gas may be reasonably applied to
non-polar gases, for example Argon (Ar). For this case, molecular dynamics
simulations were matched with experiments already in the 60s~\cite{Rahman1964}.
Corresponding parameters 
are $\sigma\approx3.4\mathring{A}$ and $\epsilon/\kB\approx120K$, which serve well
for an order-of-magnitude comparison. This would lead to a real temperature
$T^{\rm real}=T\epsilon/\kB$. In order to compare to the literature boiling point,
the density would have to be adjusted accordingly, e.g., for Argon, $T_{\rm
boiling}\approx 87.3K$~\cite{Tegeler1999} with a gas density $\rho_{\rm
boiling}\approx5.772g/l$ at atmospheric pressure.  The unit length $a$ in our
system is related to the parametrized length scale as $a= 2^{1/6}\sigma$.
Assuming a molar weight of $39.95\times1.6605\times10^{-27}kg$ for Argon, this
yields the conversion for the density $\rho^{\rm real}\approx\rho\times
1.226\times10^3g/l$. In dimensionless units, the experimental boiling
(evaporation) temperature and gas density yield $T_0\approx0.728$ and
$\rho\approx0.005$, respectively.

%
\subsection{Parallel multicanonical simulations}
The droplet condensation-evaporation transition shows a phase coexistence
between a supersaturated gas phase and a mixed phase consisting of a droplet and
a remaining gas phase. The resulting free-energy barrier~\cite{Nussbaumer2010,
Neuhaus2003} decreases the probability to pass from one phase to the other in
the canonical ensemble which leads to effects like hysteresis and other effects
characteristic for first-order like phase transitions. In order to overcome
these barriers and to sample the transition point with high accuracy, we apply
multicanonical simulations~\cite{Berg1992,JankeMUCA}. This is a natural choice for the
fixed-density scheme, while for fixed-temperature a similar approach is not
straight forward. The Boltzmann weight $\exp(-\beta E)$, where $\beta=(\kB
T)^{-1}$ with $\kB=1$, is replaced by an a priori unknown weight function which
is iteratively modified in order to yield a flat histogram~\cite{JankeMUCA}.
This allows the sampling of suppressed states between the coexisting phases and
produces accurate estimates of the observables when reweighting around the
transition temperature. As a result, we obtain a full temperature range from a
single simulation applying standard time-series and histogram reweighting
techniques.

However, this requires to sample a broad energy range with sufficiently many
tunnel events across the transition point from high energy to low energy and
vice versa. We make use of a parallel implementation of the multicanonical
method~\cite{Zierenberg2013CPC}, which allows a reduction of computational time.
In the cases of lattice gas condensation, it was tested and shown to speed up
the simulation by the number of cores used~\cite{Zierenberg2014JPCS}. We
considered lattice sizes up to $1000^2$ for the 2D and $100^3$ for
the 3D lattice gas as well as Lennard-Jones particle systems with up
to $512$ particles with a fixed density $\rho=10^{-2}$. The Monte Carlo updates
in the lattice systems include local particle shifts to a nearest-neighbor
site combined with particle displacements to a random new site. For the
Lennard-Jones gas, we restrict ourselves to local random particle displacements.  

\begin{figure*}
  \centering
  \includegraphics[]{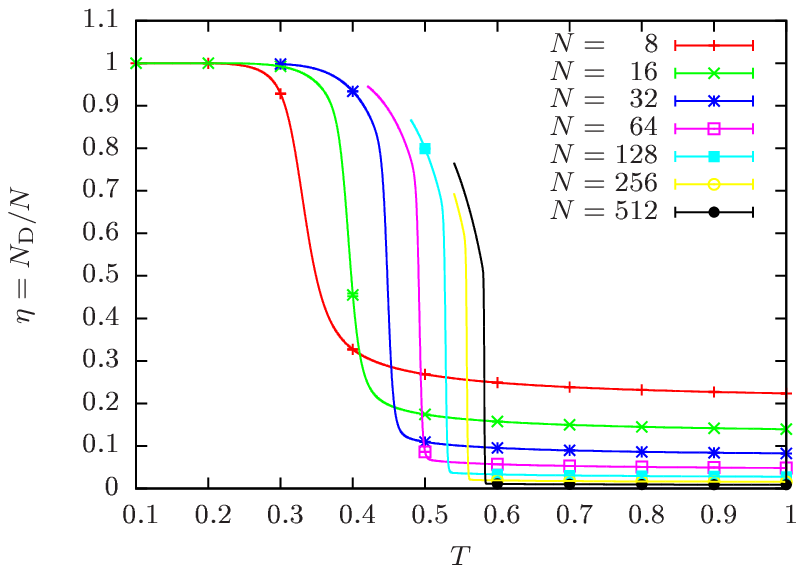}
  \hspace{2em}
  \includegraphics[]{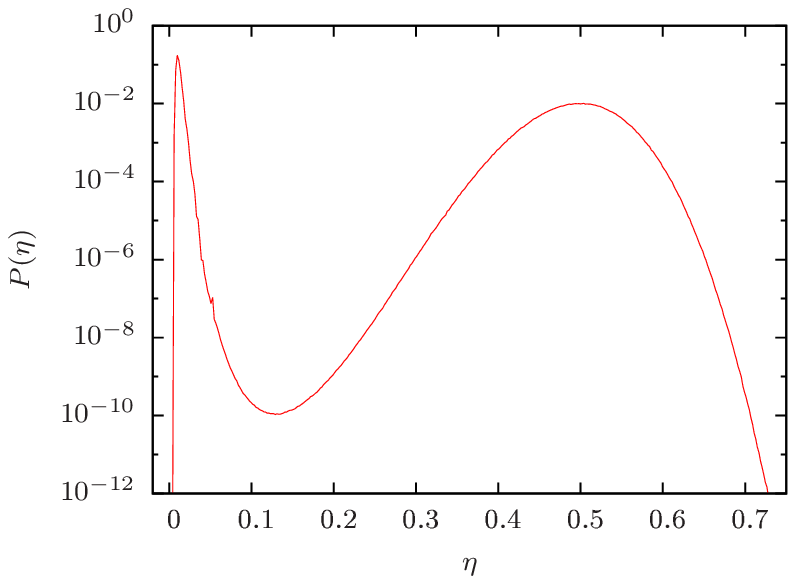}
  \caption{(Color online)
    Canonical estimates of the fraction $\eta$ of particles in the
    largest droplet for a 3D Lennard-Jones gas at fixed density $\rho=10^{-2}$~(left).
    Corresponding probability distribution of $\eta$ at $T_c$ for $N=512$~(right).
  }
  \label{fig_etaDistribution}
  \vspace{-1em}
\end{figure*}
%
\subsection{Metropolis simulations}
In addition, we applied standard Metropolis simulations in order to sample the
droplet phase at coexistence. We can use the finite-size scaling corrections 
from the fit to the transition temperatures obtained with the parallel
multicanonical simulations. Then, we make use of the first-order nature of the
transition causing a suppression of the intermediate states between the vapor
and the mixed phase. Preparing a system in the droplet phase, the Metropolis
simulation thermalizes and only samples the droplet phase for a finite time.
With increasing system size, this finite time gets larger and we may safely
sample the size of the critical droplet with a lot less statistics than would be
required for the multicanonical simulation. However, we need the multicanonical
data of suitable system sizes in order to obtain valid estimates of the
transition temperature. We do not use the Metropolis data to draw conclusions
about the transition temperature or the finite-size rounding but instead
directly measure the critical droplet size. This allows to reach an additional
order of magnitude in $N$. An example for the 3D cases is shown in
Fig.~\ref{fig_equilDroplet}.

%
\subsection{Observables}
\label{sec_observables}
The focus of this study lies on two observables and their thermal derivatives,
namely the energy $E$ and the fraction of particles in the largest cluster
$\eta=\ND/N$. The latter is determined by identifying the largest cluster as
the maximal number of connected particles $\ND$.  In case of the lattice gas,
the connected particles are simply nearest neighbors. For the Lennard-Jones gas
particles are defined as connected if $r_{ij}<2\sigma$. Thus, $\eta$ is a
measure of the ``mass'' of the largest droplet and allows one to study the
homogeneous nucleation transition efficiently. An example for the 3D
Lennard-Jones gas is given in Fig.~\ref{fig_etaDistribution}~(left).
The consideration of these types of geometric clusters is a safe choice for
dilute systems. Note that for rather dense lattice systems also the stochastic
Swendsen-Wang cluster definition has been considered~\cite{Schmitz2013}.

Anticipating a first-order phase transition, we expect in the thermodynamic
limit a discontinuity in the energy as well as in $\eta$. Hence, the thermal
derivatives of the observables $\frac{d}{dT}O = \kB\beta^2\left(\langle EO\rangle -
\langle E \rangle\langle O\rangle\right)$ will show a pronounced peak at the
transition temperature $T_c$ for finite systems, see
Fig.~\ref{fig_results_canonical}. The determination of the peak may be done very
precisely due to the full temperature range from multicanonical simulations. The
error is estimated by using jackknife error analysis~\cite{Efron1982}, where we
make use of the independent parallel production runs combining all but one time
series for each jackknife bin. In the case of the energy, the thermal derivative
is related to the specific heat $C_{\rm V}=\kB\beta^2\left(\langle E^2\rangle -
\langle E \rangle^2\right)/N$.

Moreover, we will use the mass-fraction $\eta$ in order to directly verify the
predictions about the scaling of the critical droplet size that was assumed and
used in the theory. We want to focus on the droplet phase and estimate the
average fraction of particles in the largest cluster at the transition
temperature $\critDrop$. To this end, we reweight our
multicanonical data to droplet size distributions $P(\eta)$ by time-series
reweighting, adding each reweighting factor to the histogram bin corresponding
to the observable. An example is shown in Fig.~\ref{fig_etaDistribution}~(right).
Assuming that both phases are at equal weight, we identify $\eta_{\rm min}$
such that $\tilde{Z}=\int_{\eta_{\rm min}}d\eta P(\eta)=0.5$. Then
\begin{equation}
  \critDrop = \frac{1}{\tilde{Z}}\int_{\eta_{\rm min}}d\eta\left.\eta P(\eta)\right|_{T_c}
\end{equation}
is a robust estimator for sufficiently large systems. In fact, we directly deal
with the number of particles in a droplet such that the integrals become
discrete sums. Errors are again estimated using the jackknife error analysis.
Systematic errors may arise from imperfect estimates of the coexistence
temperature, the definition of the droplet boundary and, especially for small
systems, if the distribution does not show a sufficiently pronounced dip between
the two phases.

%
\section{Results}
\label{sec_results}
 
The results are presented for a fixed particle density $\rho=N/V=10^{-2}$, up to
one exemplary case when comparing the Lennard-Jones transition temperature to
the Argon boiling point. 
This choice of a dilute system ensures that the observed transition is really
that associated with the formation of a droplet and not of a cylinder or
slap~\cite{MacDowell2006}. 
We have to emphasize in the beginning that the lattice model obviously shows
discretization effects for small systems because this explicit density may not
be realized for any system size and in general remains approximate. We will see,
however, that this does not influence our main conclusions.

We consider only a single density, because we want to test the leading-order
finite-size corrections and scaling exponents. The density will, however,
influence the finite-size scaling limit. In order to illustrate this, consider
the low-temperature series expansion of the 3D Ising model. Here, the
spontaneous magnetization is given to first order as
$m_0=1-2e^{-12\beta\Ising}+...$~. For $T\rightarrow 0$ or $\rho_0\rightarrow 0$,
this yields for the density $\rho_0=\frac{1}{2}(m_0-1)=e^{-3\beta}\left(1+ ...
\right)$ and to first order for the inverse temperature
$\beta\simeq-\frac{1}{3}\ln\rho_0$. This result may be similarly deduced from
microcanonical arguments in continuous systems (see
Ref.~\cite{Zierenberg2014JCP}).

%
\subsection{Finite-size scaling of the droplet condensation-evaporation transition temperature}
\label{sec_results_fss_temp}

\begin{figure}
  \includegraphics[]{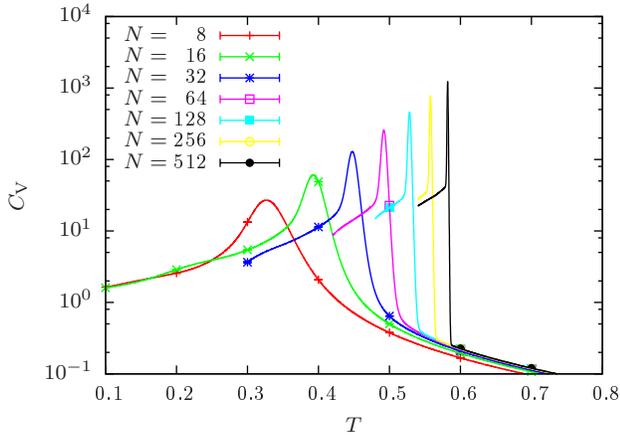}
  \caption{(Color online)
    Canonical estimates of the specific heat from parallel multicanonical
    simulations of the 3D Lennard-Jones gas at fixed density $\rho=10^{-2}$.
  }\label{fig_results_canonical}
  \vspace{-1em}
\end{figure}
%
\begin{figure}
  \includegraphics[]{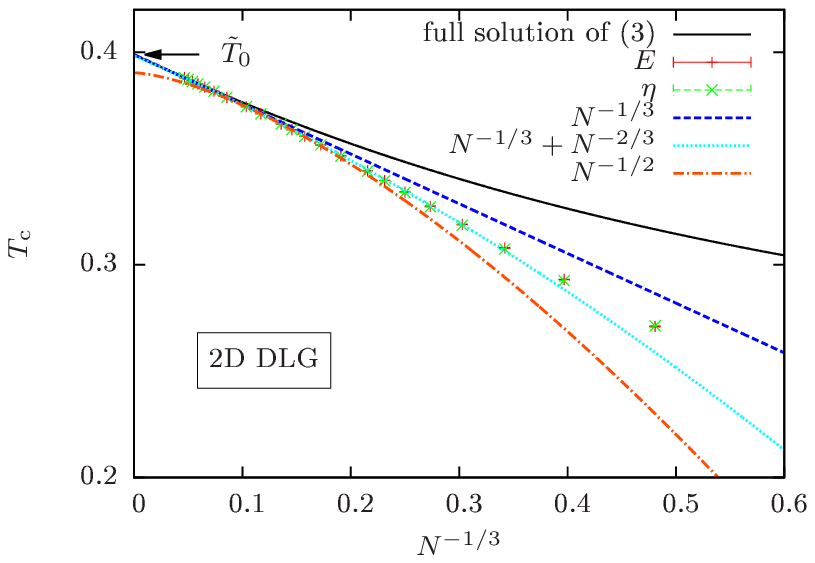}\\[2ex]
  \includegraphics[]{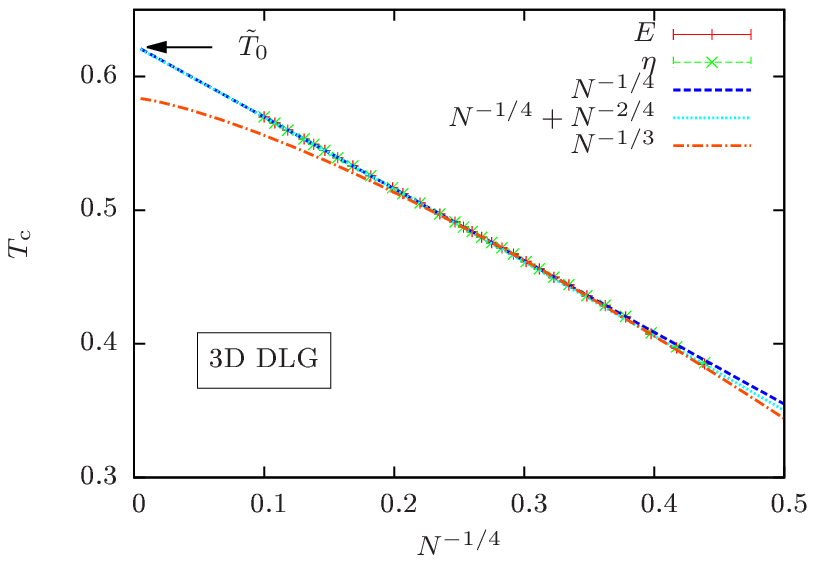}\\[2ex]
  \includegraphics[]{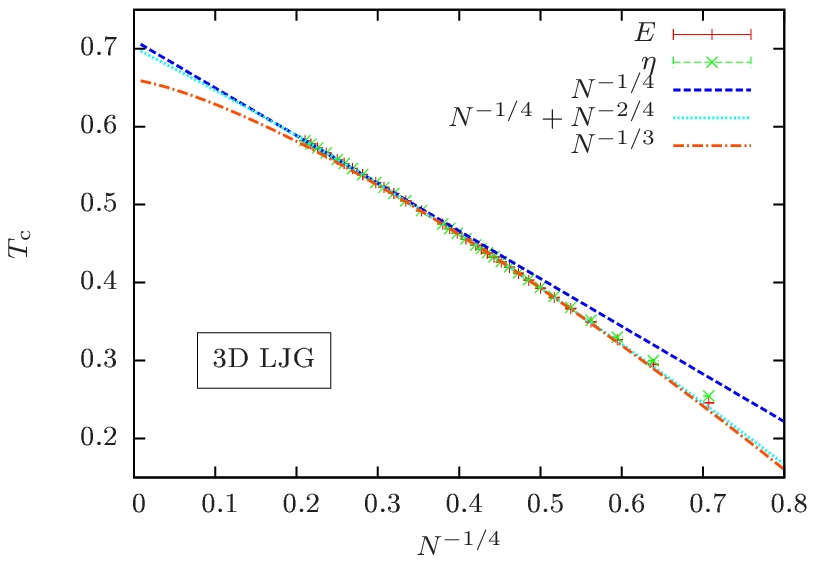}\\[1ex]
  \caption{
    (Color online)
    Finite-size scaling of the droplet condensation-evaporation transition
    temperature for the 2D lattice, 3D lattice and 3D Lennard-Jones gas from
    top to bottom. The lattice gas is compared to theoretical predictions (see
    text and also Sec.~\ref{sec_fss_temp}).
  }\label{fig_results_fss_temp}
  \vspace{-1em}
\end{figure}

The transition temperature, where the transition-related observables show a
discontinuity, may be estimated by the peak locations of their thermal
derivatives, for an example see Fig.~\ref{fig_results_canonical}.  In the limit
of large systems, this transition temperature is predicted to scale as 
\begin{equation}
  T_c = T_0 + aN^{-\frac{1}{d+1}} + ... , 
\end{equation}
see Sec.~\ref{sec_fss_temp}. Figure~\ref{fig_results_fss_temp} shows numerical
results for the three considered situations, the lattice gas in two and three
dimensions and the Lennard-Jones gas in three dimensions. The estimated
transition temperatures are obtained from the maxima of the specific heat (red
pluses) as well as the thermal derivative of the fraction of particles in the
largest cluster $\frac{d\eta}{dT}$ (green crosses). Both estimates are
remarkably similar as expected for first-order phase transitions, and hence we
only show local fits to the estimated transition temperature derived from the
specific heat.

\newcolumntype{C}[1]{>{\centering\let\newline\\\arraybackslash\hspace{0pt}}m{#1}}
\renewcommand\arraystretch{1.3} 
\newcommand{\cw}{5em}
\begin{table*}
  \centering
  \caption{\label{tab_results_fss_temp}%
    Results of different fit functions to the condensation-evaporation
    transition temperature with the statistical errors of the fits. If no upper
    range is provided, it refers to $N_{\rm max}=10\,000$ for the lattice gas
    cases (2D and 3D) and $N_{\rm max}=512$ for the Lennard-Jones gas. The
    reference infinite-size value $\tilde{T}_0$ is obtained for the lattice
    systems from the Onsager solution (2D)~\cite{Onsager} and from
    low-temperature series expansions (3D)~\cite{3DLowTempM}. 
  }
  \begin{tabular}{ C{\cw}|C{\cw} C{\cw} C{3em}|C{\cw} C{\cw} C{3em}|C{\cw} C{\cw} C{3em}|C{\cw} }
    \toprule
    Model  & \multicolumn{3}{c|}{$T_0+aN^{-1/(d+1)}$} &
    \multicolumn{3}{c|}{$T_0+aN^{-1/(d+1)}+bN^{-2/(d+1)}$} &
                     \multicolumn{3}{c|}{$T_0+aN^{-1/d}$} & Ref\\
    & Range & $T_0$ & $\chi^2$ & Range & $T_0$ & $\chi^2$ & Range &  $T_0$ &
    $\chi^2$ & $\tilde{T}_0$ \\\colrule
      2D DLG & [\makebox[2.0em][r]{2500}:\makebox[1.6em][r]{   }] & 0.39884(3) & 0.4 & 
               [\makebox[2.0em][r]{ 400}:\makebox[1.6em][r]{   }] & 0.3982(1)  & 1.8 &
               [\makebox[1.6em][r]{ 324}:\makebox[1.6em][r]{900}] & 0.3903(2)  & 8.0 & $0.39882\dots$  \\
      3D DLG & [\makebox[2.0em][r]{2160}:\makebox[1.6em][r]{   }] & 0.62341(4) & 0.6 & 
               [\makebox[2.0em][r]{1663}:\makebox[1.6em][r]{   }] & 0.6229(4)  & 1.9 &  
               [\makebox[1.6em][r]{  68}:\makebox[1.6em][r]{243}] & 0.5840(3)  & 1.6 & $\approx 0.622$ \\
      3D LJG & [\makebox[2.0em][r]{ 160}:\makebox[1.6em][r]{   }] & 0.7106(4)  & 0.8 & 
               [\makebox[2.0em][r]{  10}:\makebox[1.6em][r]{   }] & 0.7011(4)  & 1.1 &  
               [\makebox[1.6em][r]{  12}:\makebox[1.6em][r]{ 48}] & 0.6597(4)  & 1.0 &                 \\
    \botrule
  \end{tabular}
\end{table*}

In all three cases, we observe for large system sizes a proper finite-size
scaling behavior according to the predicted scaling, shown by a good
quality of the least-squares fits
(a reduced $\chi^2$ per degree of freedom of about $1$). In accordance
with the literature, we verify that this requires rather large system sizes for
the leading order fit (dashed dark blue fit). The intermediate sized systems
may be included in an empirical fit with the next higher orders (dotted light
blue fit), i.e., \mbox{$T_c = T_0 + aN^{-1/(d+1)}+bN^{-2/(d+1)}$}. In addition,
we want to mention that for intermediate system sizes an effective different
scaling behavior may be observed. While not completely apparent, it may be
justified by sufficiently good $\chi^2$, that the transition temperature can be
locally describes by a $N^{-1/d}$ behavior (dash-dotted orange fit). For
details of the individual fits see Table~\ref{tab_results_fss_temp} and the
following discussion. The scaling of the intermediate regime is consistent with
studies of flexible homopolymer aggregation~\cite{Zierenberg2014JCP}, where a
large fraction of the system is involved in the formation of the
droplet/aggregate. If almost all constituents are involved in the transition,
then the linear extension of the homogeneous, isotropic condensate is just
$N^{1/d}$ -- which justifies the assumed scaling behavior. This leads to two
possible conclusions: Either the local scaling function is a polymer property
or it is a generic property for a small number of polymers, where the latter
seems to be in agreement with the presented results.  In order to explore this
observation further, we will investigate the size of the critical droplet and
the rounding of the transition (which in fact provides suitable fitting ranges
for the transition temperature) in the following subsections. 

In the case of the 2D lattice gas, we may compare directly to the analytic
solution of the infinite system and to the full solution of
Eq.~\eqref{eq_koteckySizeDep} for finite systems derived in
Sec.~\ref{sec_fss_temp}. The full solution shows large deviations for small
systems as expected. With increasing system size, however, it starts to describe
the finite-size scaling approximately. This already gives a hint to the choice
of a proper fitting-range with a leading-order scaling behavior. A least-square
fit of the leading-order for the largest system sizes $N\ge2500$ yields an
adequate $\chi^2\approx0.4$ and an infinite-size transition temperature
\mbox{$T_0=0.39884(3)$}. This is consistent with the analytic result inverting
Onsager's solution for the magnetization~\cite{Onsager}, with $T=\TI/4$, 
\begin{equation}
  m_0(\tilde{T}_0)= 0.98 = \left[ 1- \sinh^{-4}(1/2\tilde{T}_0)\right]^{1/8}. 
\end{equation}
The resulting \mbox{$\tilde{T}_0=0.39882$} is shown in
Fig.~\ref{fig_results_fss_temp}~(top) by the arrow. Including the next order and
fitting $N\ge400$ yields a $\chi^2\approx1.8$ with $T_0=0.3982(1)$, which
slightly deviates from the exact solution. This may be taken as a hint that our
next-order term is only an effective correction and additional corrections of
the same order may be apparent. Interesting to compare is also the amplitude of
the leading-order correction $aN^{-1/3}$. The linear and higher order fits yield
$a=-0.234(1)$ and $a=-0.214(2)$ respectively. This may be compared to the
power-series expansion of the full solution (see
Eq.~\eqref{eq_kotecky_expansion} and Eq.~\eqref{eq_koteckyIsing}). Making use of
the analytic solution for $m_0$, the series expansion for $\chi$ up to $300^{\rm
th}$ order and the integral solution of $\tauW$ (for a list of equations see
Ref.~\cite{Nussbaumer2008}), we may numerically differentiate
Eq.~\eqref{eq_koteckyIsing} and calculate in lattice gas units
$a=\Delta_c^{2/3}\rho^{1/3}/4f'(m,\TI_0)\approx-0.239$ in decent agreement with
the leading-order fit. The previously mentioned intermediate scaling regime is
not very prominent for the 2D lattice gas where a least-square fit
to $N^{-1/2}$ in the (already small) range $N=[324, 900]$ still yields a
$\chi^2\approx 8$. Moreover, the infinite-size extrapolation is obviously wrong.
It is worth noting, that the intermediate scaling regime in polymer
aggregation~\cite{Zierenberg2014JCP} was observed in three dimensions, which
suggests that the prominence of this regime may depend on the dimension.

For the 3D lattice gas, we may compare the finite-size scaling results of the
infinite-size transition temperature to low-temperature expansions, see
Sec.~\ref{sec_fss_temp}. A fit of the leading-order scaling behavior to
sufficiently large systems in the range $N\ge2160$ yields $T_0=0.62341(4)$
with $\chi^2\approx0.6$ which is in the vicinity of the (not exact)
low-temperature expansion $\tilde{T}_0\approx0.622$~\cite{3DLowTempM}, again
shown by the arrow. Increasing the fit range still yields reasonably good fits
with higher $\chi^2$ of the same quality as the fit including the next order
$N^{-2/4}$, see also Table~\ref{tab_results_fss_temp}. Notice that in the case
of the 3D lattice gas also the small system sizes seem to coincide with the
leading order fit. However, considering only an intermediate regime allows to
fit the $N^{-1/3}$ behavior with qualitatively good local agreement. If the
larger system sizes were not present, this could be interpreted as the leading
order scaling behavior. Comparing the fit to this effective ansatz with the
low-temperature series expansion shows, however, strong deviations. 

In the case of the 3D Lennard-Jones gas one may best see the arising
peculiarities.  The leading-order fit for $N\ge160$ to $N^{-1/4}$ yields
$T_0=0.7106(4)$ with $\chi^2\approx0.8$ but shows a clear deviation for small
system sizes. Including the next order for $N\ge10$ yields $T_0=0.7011(4)$ with
$\chi^2\approx1.1$, in rough agreement with the leading-order result, and
recaptures the deviation of the small system sizes. This is consistent with
results for the same Lennard-Jones model~\cite{Trokhymchuk1999,Chapela1977}.
However, considering an intermediate regime $N=[12:48]$ with the ansatz
$N^{-1/3}$ yields a qualitatively good fit with $T_0=0.6597(4)$ and
$\chi^2\approx 1$, which deviates strongly from the $N^{-1/4}$ fit. Again, if
the largest system sizes were not present this could be interpreted as the
finite-size scaling corrections, especially if no reference temperature is
available. Locally it seems that this is an intermediate scaling regime which
is, however, already covered by the theoretically predicted scaling behavior
including empirical higher-order corrections. This shows the necessity to
control that the available data is really in the expected leading-order scaling
regime. 

Comparing to the boiling temperature of Argon, we have to exemplary consider a
density $\rho=5\times10^{-3}$. Then, the leading-order fit yields
$T_0=0.6525(3)$ (not shown here) which differs from the experimental result
$\tilde{T}_0\approx0.728$ (see Sec.~\ref{sec_model_LJ}) by $\sim10\%$. While the
order-of-magnitude is comparable, the difference is expected for the truncated
Lennard-Jones potential~\cite{Trokhymchuk1999}.

The largest systems considered for the Lennard-Jones gas included 512 particles
in a box of length $L\approx 37.1\approx41.7\sigma$. This is a lot smaller than
the system sizes considered in Ref.~\cite{Binder2009} ($L\le100\sigma$) at fixed
temperature $T\approx0.68$ (for their parameterization) with typical particle
numbers $N\approx15\,800$. Still, in this fixed-$T$ approach they did not see
the predicted scaling behavior of the transition density shift but needed to
extrapolate an effective exponent (smaller than $-0.89$) in order to recover the
theoretical prediction $L^{-0.75}$. Similarly, a direct fit of the leading-order
power-low exponent to the present data yields $N^{-0.28(1)}\propto L^{-0.84(2)}$
already for smaller system sizes $N\ge160$.  This implies that an orthogonal
phase boundary crossing may lead in certain situations to reduced finite-size
corrections and serves as a useful, complementary approach. 

\vspace{-2mm}
%
\subsection{Finite-size rounding of the droplet condensation-evaporation transition}
\begin{figure}
  \includegraphics[]{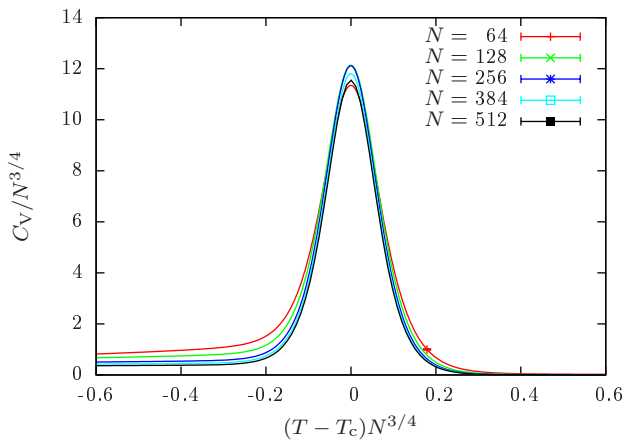}
  \caption{(Color online)
    Example of the transition rounding for the specific heat of the 3D
    Lennard-Jones gas. The $x$ and $y$ axes are rescaled according to the
    leading-order scaling behavior.
  }
  \label{fig_results_rounding}
  \vspace{-1em}
\end{figure}
\begin{figure}
  \includegraphics[]{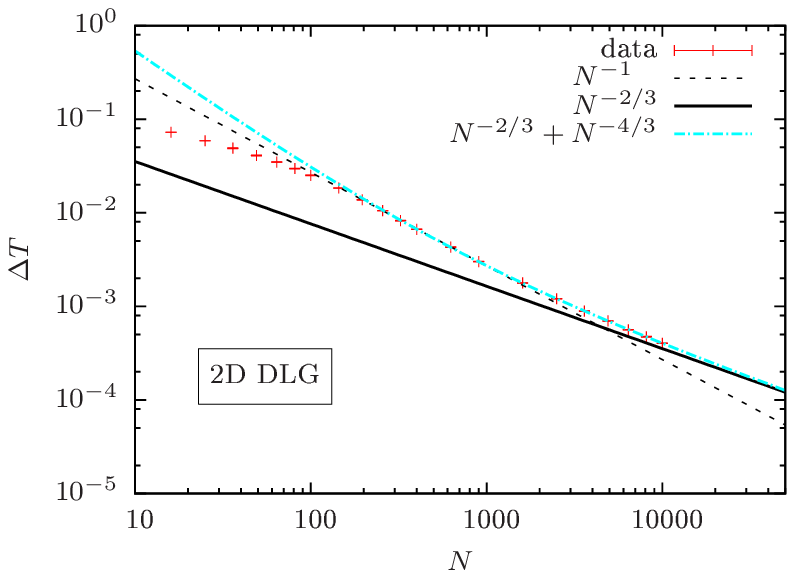}\\[2ex]
  \includegraphics[]{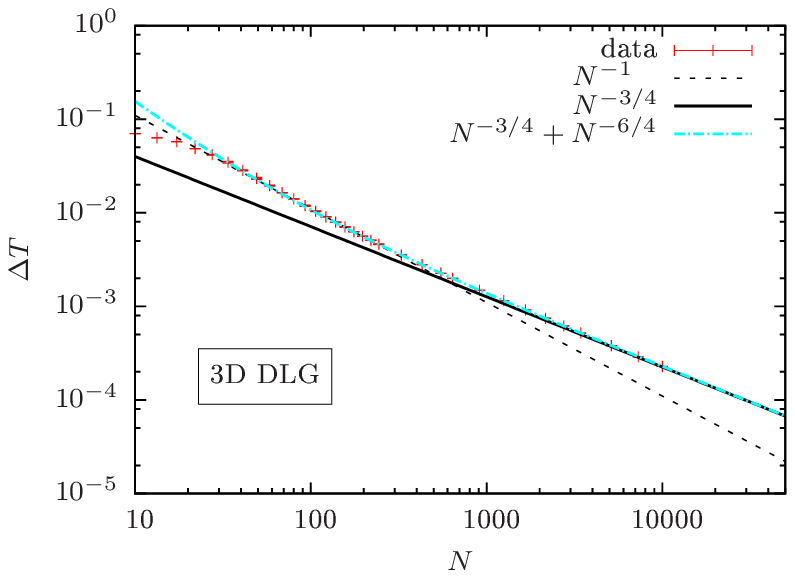}\\[2ex]
  \includegraphics[]{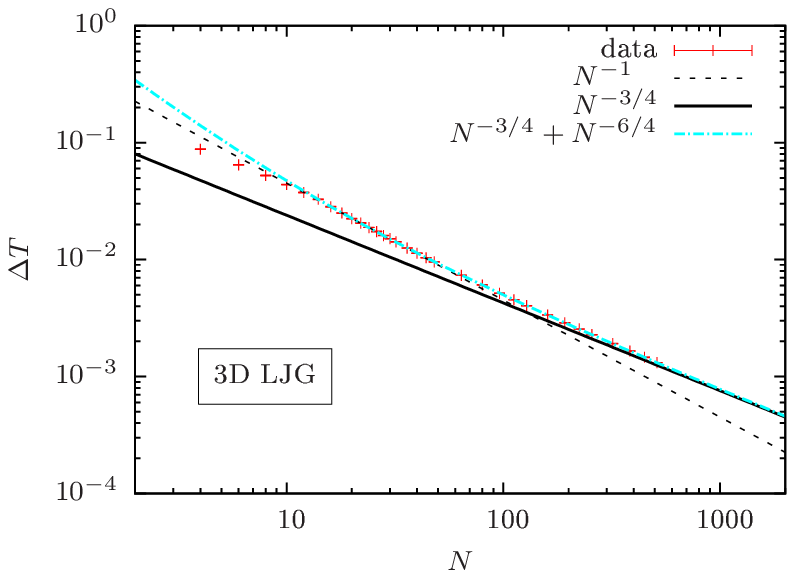}\\[1ex]
  \caption{(Color online)
    Finite-size rounding of the droplet condensation-evaporation transition for
    the 2D lattice, 3D lattice and 3D Lennard-Jones gas from top to bottom. The
    results are presented for $\rho=0.01$ but are consistent for different
    densities.  
  }\label{fig_results_fss_rounding}
  \vspace{-1em}
\end{figure}
In the limit of large systems, the rounding of the transition is predicted to
scale as $\Delta T\propto N^{-d/(d+1)}$ for fixed density, see
Sec.~\ref{sec_fss_rounding}.  Figure~\ref{fig_results_rounding} shows the
rescaled specific heat from Fig.~\ref{fig_results_canonical} for the 3D
Lennard-Jones gas. At a first-order transition, the maximum of the specific
heat should scale inverse proportional to the rounding of the
transition~\cite{Imry1980}. The rounding of the transition is here estimated as
the half-width of the specific-heat peak, defined as the width where $C_{\rm
V}\ge\frac{1}{2}C_{\rm V}^{\rm max}$. Errors are obtained by jackknife error
analysis.

Figure~\ref{fig_results_fss_rounding} shows the finite-size scaling of the
rounding for all three considered systems.  In all cases, we could clearly
identify two different scaling regimes: For intermediate system sizes one may
observe $\Delta T \propto N^{-1}$ (dashed line) and for large system sizes
$\Delta T \propto N^{-d/(d+1)}$ (solid line). The latter is the expected
scaling of the rounding as predicted by theory~\cite{Binder2003,Binder2009} and
reformulated in Sec.~\ref{sec_fss_rounding}. Thus, having only data available in
the intermediate regime would suggest a wrong finite-size scaling behavior
consistent with the intermediate regime for the transition temperature.
However, the intermediate regime is nicely fitted by considering the predicted
scaling behavior for large system sizes and empirically including the next
higher-order corrections (dash-dotted light blue fit), i.e., $\Delta T =
a'N^{-d/(d+1)}+b'N^{-2d/(d+1)}$.

In the case of the 2D lattice gas, smallest systems show no systematic behavior
which explains the large deviations in Fig.~\ref{fig_results_fss_temp}. For the
intermediate regime one can see a direct particle or volume dependence ($\rho$
fixed). The onset of the large-system regime is consistent with the scaling
range in the transition temperature that approximately coincides with the full
solution (Fig.~\ref{fig_results_fss_temp}). A fit including higher order
corrections yields a $\chi^2\approx0.7$ including already system sizes
$N\ge324$ and thus including the intermediate regime. 

For the 3D lattice gas and the Lennard-Jones gas, the intermediate regime is
apparent for quite small systems and also shows the $N$-proportionality up to
the crossover to the large-system regime. Again, the crossover is consistent
with a good choice of a leading-order fitting range for the finite-size scaling
of the transition temperature. The fit to the rounding of the transition,
including higher-order corrections, allows to include system sizes $N\ge175$
with $\chi^2\approx2.3$ for the lattice gas and $N\ge16$ with
$\chi^2\approx0.9$ for the Lennard-Jones gas case. Again, this includes (parts
of) the intermediate regime by considering empirical higher-order corrections.

Previous studies of the lattice gas in two and three dimensions at fixed-$\rho$
showed significant deviations from the predicted exponents for the transition
rounding~\cite{Malakis2007}, using average densities of states from Wang-Landau
simulations. For the 3D case they found an effective scaling of the rounding with
$L^{-2.45(2)}\propto N^{-0.82(1)}$. The present results on the other hand
clearly confirm the ``large''-system scaling behavior. Direct fits of the
power-law behavior to the largest system sizes of the 3D lattice gas model yield
effective exponents $N^{-0.78(1)}$ and $N^{-0.76(1)}$ for $N\ge1663$ and
$N\ge5120$ respectively, close to the predicted scaling $N^{-0.75}$. 

The finite-size rounding shows to be a good observable to identify the
previously noticed intermediate regime. The width of the transition is
associated to the fluctuations of the system~\cite{Imry1980}, which should
depend on the inverse volume of the relevant system size. The relevant system
size remains the droplet, which scales for large systems as $R^d\propto
N^{d/(d+1)}$. On the other hand, for small systems the droplet includes a large
fraction of the system, see the following discussion. Thus the local scaling of
the intermediate regime is not surprising. Knowing the scaling behavior for
large systems, however, allows to anticipate this gradual crossover with
empirical higher-order corrections. 

%
\subsection{Scaling regimes of the critical droplet size}

We identified the leading-order finite-size scaling corrections as powers of the
critical droplet radius. This requires that the mass of the critical droplet
scales in the limit of large systems as $N^{d/(d+1)}$. We want to test this
relation by considering the fraction of particles in the largest cluster $\eta$
--- or in other words the mass of the droplet divided by the total number of
particles --- that should scale as $N^{-1/(d+1)}$. 

\begin{figure}[t!]
  \includegraphics[]{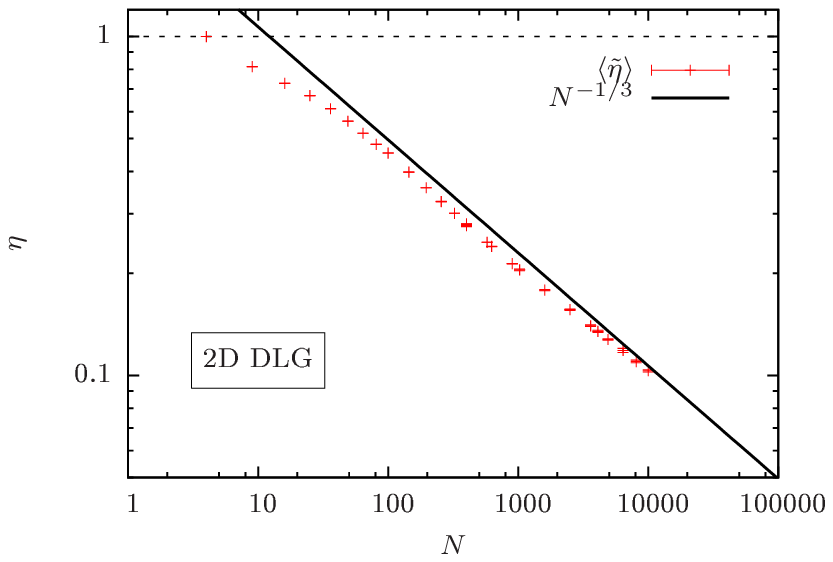}\\[2ex]
  \includegraphics[]{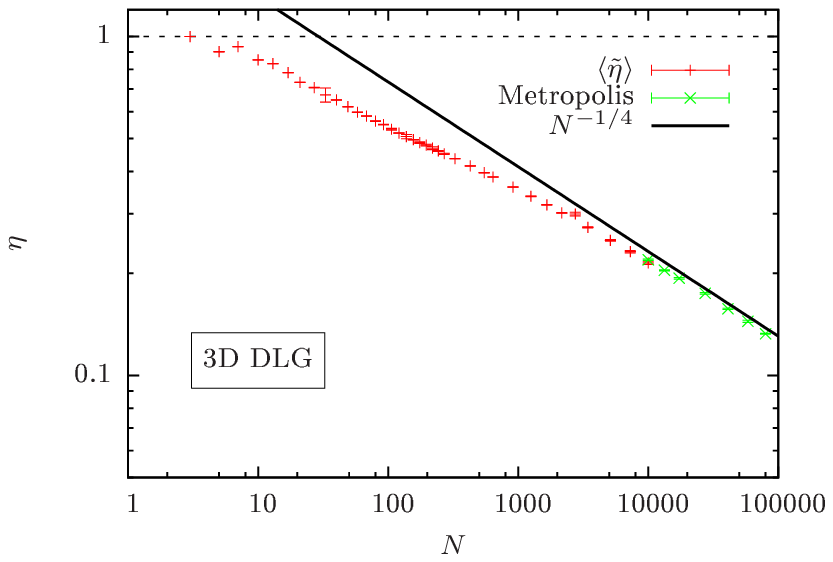}\\[2ex]
  \includegraphics[]{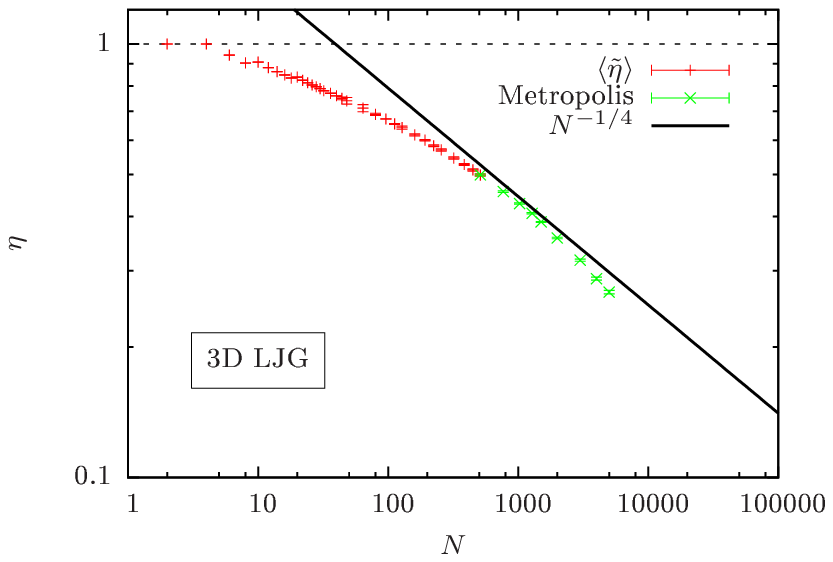}\\[2ex]
  \caption{(Color online)
    Scaling of the particle fraction in the largest droplet $\eta$ for 2D
    lattice, 3D lattice and 3D Lennard-Jones gas from top to bottom. For large
    systems, $\eta$ should scale as $N^{-1/(d+1)}$. For small system sizes, the
    majority of particles is in the droplet $\eta=\mathcal{O}(1)$ (horizontal
    dashed line).  The Metropolis data are obtained in the droplet phase at the
    finite-size transition temperature as extrapolated from the linear fit in
    Fig.~\ref{fig_results_fss_temp}.
  }\label{fig_results_criticalDroplet}
  \vspace{-1em}
\end{figure}
Figure~\ref{fig_results_criticalDroplet} shows that this assumption seems to be
valid for all three considered models. In each case, we show the expectation
value $\critDrop$ in the droplet phase at the transition temperature. The
horizontal dashed line at $\eta=1$ corresponds to all particles being in the
droplet. For one, the results from our multicanonical simulation (red pluses)
are measured as described in Sec.~\ref{sec_observables} with jackknife errors.
In addition, there are statistical averages from canonical Metropolis
simulations (green crosses) prepared in the droplet phase at
$T_c(N)=T_0+aN^{-1/(d+1)}$ as extrapolated in Sec.~\ref{sec_results_fss_temp}.
To be precise, we considered the following pairs $(T_0,a)$: $(0.623,-0.537)$
for the 3D lattice gas as well as $(0.710,-0.611)$ for the 3D Lennard-Jones
gas. In this case, the errors are statistical errors including the integrated
autocorrelation times. The straight line is the expected scaling behavior
shifted as a guide to the eye in order to be compared with the numerical data. 

The 2D lattice gas shows the expected scaling behavior quite early, already for
roughly 100 particles. In the case of the 3D lattice gas the situation already
changes and the expected scaling behavior only starts for quite large system
sizes of the order of 2000 particles. This is consistent with the 3D
Lennard-Jones gas, where, however, the available system sizes are much smaller.
Still the expected scaling may be anticipated. Including the Metropolis data,
it shows that the leading-order extrapolation of the available transition
temperatures is not precise enough and seems to overestimate the transition
temperature for larger systems, consistent with the observations in
Sec.~\ref{sec_results_fss_temp}. 

All cases show that for small system sizes, the majority of particles is
included in the single macroscopic droplet. A corresponding constant behavior
$\critDrop\sim1$ (the horizontal dashed line) would explain an intermediate scaling
regime where the relevant length scale depends on the system volume. From
Fig.~\ref{fig_results_criticalDroplet}, this may be at most guessed locally.
However, regarding the measurement of the average critical droplet we have to
mention that the small systems including only a few particles also show narrow
distributions. These are discretized by nature and only defined in the range
$[1,N]$. Thus, the observable may not provide the required measurement
sensitivity. Still, having a large fraction of particles in the droplet,
especially compared to the vanishing fraction in the limit of large systems,
suffices for the argument that locally an effective intermediate scaling becomes
visible. Moreover, it becomes clear that there are corrections to the size of
the critical droplet for small and intermediate systems. This should also have
an effect on the scaling of the transition temperature and rounding.

\section{Conclusions}
\label{sec_conclusions}

We presented results on the canonical finite-size scaling of the droplet
condensation-evaporation transition. To this end, we performed a comparative
study in which we considered two- and three-dimensional lattice gas as well as
three-dimensional off-lattice Lennard-Jones gas models in order to draw general
conclusions. In all cases, we could verify the theoretical predictions about the
finite-size corrections to the transition temperature $\propto N^{-1/(d+1)}$ and
the finite-size rounding $\propto N^{-d/(d+1)}$ in the limit of large
systems~\cite{Biskup2002,Binder2003}. In addition, we identified an intermediate
regime with an effective different scaling behavior in all three cases, most
apparent in the finite-size rounding. This regime is consistent with results
from flexible polymers~\cite{Zierenberg2014JCP}, where the aggregate included
most of the polymers. Our measurement of the average droplet size in equilibrium
with vapor at coexistence showed that indeed for small system sizes a large
fraction of the system is included in the single macroscopic droplet. The
scaling of the droplet size, however, is nicely described by the theoretical
predictions~\cite{Binder1980, Binder2003} in the limit of large systems. 

Without knowledge of the larger system sizes, the intermediate regime could give
way to a wrong finite-size scaling behavior with qualitatively good fit results
for both the transition temperature and rounding. However, we were able to show
that there is a gradual crossover to the large-system scaling regime which can
be approximately described by including empirical higher-order corrections to the
theoretical scaling predictions, namely polynomial orders of the critical
droplet size. If only the leading-order corrections are considered, the rounding
is a good source getting a lower fit limit for a qualitatively good fit. This
was compared to analytic and low-temperature results for the lattice cases. It
is thus also suitable in order to identify the finite-size scaling regime for
phase transitions with mixed phases, like the condensation-evaporation
transition but also more complex systems like polymer aggregation. Considering
only the local, intermediate regime and applying standard finite-size scaling
approaches (with possibly wrong corrections) would then lead to wrong estimates
of the infinite-system limit. A similar note of caution has been recently
demonstrated in the finite-size scaling of self-avoiding walks on percolation
clusters~\cite{Fricke2014}.

An intuitive approach to the leading-order finite-size scaling corrections is
the competition of volume ($L^d$) and surface ($L^{d-1}$) contributions. For a
first-order phase transition, this gives rise to a finite-size correction of the
order $L^{-1}$, where $L$ is the relevant length scale of the
system~\cite{Privman1990, Borgs1995, Borgs2002, Zierenberg2014JCP}. We argue
that the linear extension of the droplet $R$ at coexistence plays this dominant
role. The condensation-evaporation transition clearly connects a homogeneous and
a mixed phase in the canonical ensemble.  However, within the canonical
approach, we may consider a (virtual) subsystem with the volume of the
transition droplet. By translational invariance, this subsystem may be always
constructed around the largest droplet. Above the transition temperature, this
subsystem includes a homogeneous gas phase while at and below the transition it
is filled by the largest droplet and hence shows a homogeneous liquid phase.
Thus, this may be interpreted as a grand-canonical transition between
homogeneous phases in the virtual system spanned by the volume of the critical
droplet. By construction, this virtual volume would have open boundary
conditions yielding a finite-size shift of order $R^{-1}$ and a finite-size
rounding of order $R^{-d}$. This picture is consistent with rigorous results for
non-periodic first-order phase transitions~\cite{Borgs2002}. However, this
argument relies on the finite-size scaling of $R\propto N^{1/(d+1)}$, which was
shown to be already non-trivial.

The present study shows that considering an orthogonal crossing of the phase
boundary still yields the same finite-size corrections and serves as a
complementary tool. On the example of the transition rounding, the fixed-density
approach was shown to be closer to the expected large-system scaling behavior
already for smaller system sizes. In general, both directions have their
advantages and drawbacks both numerically and systematically. This may be
exploited for one's benefit by choosing the suitable direction for the problem
at hand, as was also recently demonstrated for the Blume-Capel
model~\cite{Zierenberg2015}.

\begin{acknowledgments}
We thank A. Malakis, M. Mueller and P. Schierz for useful discussions.
The project was funded by the European Union and the Free State of Saxony.
Computing time provided by the John von Neumann Institute for Computing (NIC)
under grant No.~HLZ21 on the supercomputer JUROPA at J\"ulich Supercomputing
Centre (JSC) is gratefully acknowledged.
Part of this work has been financially supported by 
 the Deutsche Forschungsgemeinschaft (DFG) 
   through the Leipzig Graduate School of Excellence GSC185 ``BuildMoNa''
 and by the Deutsch-Franz\"osische Hochschule (DFH-UFA) 
   through the German-French Graduate School under Grant No.\ CDFA-02-07. 
\end{acknowledgments}

\bibliographystyle{model1-num-names}

\end{document}